\documentclass{aa}
\usepackage{upgreek}
\usepackage{graphicx}
\usepackage[varg]{txfonts}
\usepackage{natbib}
\usepackage{soul}
\newlength{\linwx}
\usepackage{color}

\begin{document}

\title{Rocky super-Earths or waterworlds: the interplay of planet migration, pebble accretion and disc evolution}
\author{
Bertram Bitsch \inst{1},
Sean N. Raymond \inst{2},
\and
Andre Izidoro \inst{3}
}
\offprints{B. Bitsch,\\ \email{bitsch@mpia.de}}
\institute{
Max-Planck-Institut f\"ur Astronomie, K\"onigstuhl 17, 69117 Heidelberg, Germany
\and
Laboratoire d'Astrophysique de Bordeaux, CNRS and Universit\'e de Bordeaux, All\'ee Geoffroy St. Hilaire, 33165 Pessac, France
\and
UNESP, Univ. Estadual Paulista - Grupo de Din\`amica Orbital Planetologia, Guaratinguet\`a, CEP 12.516-410, S\~ao Paulo, Brazil
}
\abstract{
Recent observations have found a valley in the size distribution of close-in super-Earths that is interpreted as a signpost that close-in super-Earths are mostly rocky in composition. However, new models predict that planetesimals should first form at the water ice line such that close-in planets are expected to have a significant water ice component. Here we investigate the water contents of super-Earths by studying the interplay between pebble accretion, planet migration and disc evolution. Planets' compositions are determined by their position relative to different condensation fronts (ice lines) throughout their growth. Migration plays a key role. Assuming that planetesimals start at or exterior to the water ice line ($r>r_{\rm H_2O}$), inward migration causes planets to leave the source region of icy pebbles and therefore to have lower final water contents than in discs with either outward migration or no migration. The water ice line itself moves inward as the disc evolves, and delivers water as it sweeps across planets that formed dry. The relative speed and direction of planet migration and inward drift of the water ice line is thus central in determining planets' water contents. If planet formation starts at the water ice line, this implies that hot close-in super-Earths (r<0.3 AU) with water contents of a few percent are a signpost of inward planet migration during the early gas phase. Hot super-Earths with larger water ice contents on the other hand, experienced outward migration at the water ice line and only migrated inwards after their formation was complete either because they become too massive to be contained in the region of outward migration or in chains of resonant planets. Measuring the water ice content of hot super-Earths may thus constrain their migration history. 
}
\authorrunning{Bitsch et al.}\maketitle

\section{Introduction}
\label{sec:Introduction}

The most common type of planet around other stars are close-in super-Earths with orbital periods of less than 100 days \citep{2011arXiv1109.2497M, 2018AJ....156...24M}. These planets have roughly radii of 1-4 Earth radii and weight up to 20 Earth masses. Additionally most of these detected planetary systems are in multi-planet systems with low mutual inclinations and low eccentricities \citep{2011arXiv1109.2497M, 2011ApJS..197....1M, 2016PNAS..11311431X, 2018arXiv180700549V}. The formation pathways of these systems is still far from being completely understood.

The composition of the planet can hold important clues to its formation pathway. If the mass of a planet is known through RV detections and its radius through transit observations, the mean density of it can be calculated. This gives important information about the planetary composition through interior structure models \citep{2007ApJ...665.1413V, 2007Icar..191..337S, 2007ApJ...669.1279S, 2007ApJ...659.1661F, 2007Icar..191..453S, 2008ApJ...673.1160A, 2013PASP..125..227Z, 2016AJ....152..160B}. A planet with a mean density consistent with terrestrial planets most likely formed in the inner regions of the disc without significant accretion of water ice, while planets with lower densities could harbour a significant fraction of water ice. On the other hand, the observed planetary radius can also be greatly influenced by its atmosphere.

Recent analysis of the Kepler data with follow up analysis of the host-stars have revealed a gap in the radius distribution of these super-Earths planets \citep{2017AJ....154..109F}. This divides the super-Earths in two populations, one with a peak at 1.3-1.5 Earth radii and another population with a peak at 2.3-2.5 Earth radii, separated by a gap in the distribution at about 1.8 Earth radii.

Considering that these planets are sufficiently close to the their host star, photoevaporation of the planetary atmosphere of small planets can destroy their entire atmospheres leaving the bare planetary cores behind \citep{2012MNRAS.425.2931O, 2013ApJ...775..105O, 2017ApJ...847...29O, 2018ApJ...853..163J}. The same process could even also evaporate atmospheres of hot Jupiters \citep{2004A&A...419L..13B}, even though this is under debate \citep{2007ApJ...658L..59H, 2007Icar..187..358H}.

Using the assumption that the observed close in super-Earths are bare cores and that thus the measured planetary radius corresponds to the core radius, interior models of planets have revealed that the planetary radius distribution is more consistent with rocky planets than with icy planets. 

In photoevaporation models, a rocky composition produces the gap in planetary radii at 1.8 Earth radii, which seems inconsistent with mostly icy super-Earths \citep{2013ApJ...775..105O, 2014ApJ...792....1L, 2017ApJ...847...29O, 2018ApJ...853..163J}. On the other hand, \citet{2014A&A...562A..80K} find that all water ice on the planet could be evaporated away by the host star, leaving a rocky core behind independently if water ice was originally present in the planet or not, if the planet is below 3 Earth masses and very close to the host star ($r_{\rm p}<$0.03AU).

However, recent more detailed measurements of stellar distances and stellar radii seem to have filled in the gap at 1.8 Earth radii to some level \citep{2018arXiv180501453F}, indicating that some fraction of super-Earth planets could indeed be water rich. Taking the effect of binary stars onto the observations of planetary radii into account, \citet{2018arXiv180410170T} have shown that the radius gap could be reduced even more. Independent observations by \citet{2018MNRAS.479.4786V} find that the gap in the planetary radius distribution has shifted to 2.0 Earth radii, also more consistent with water rich super-Earths. 

Another alternative to explain the valley in the radius distribution was put forward by \citet{2018arXiv181103202G}. They show that the planetary evolution based on a core-powered mass-loss mechanism alone can produce the observed valley in the radius distribution. \citet{2018arXiv181103202G} also find in their model that super-Earths should be predominately rocky, but can contain up to 20\% water. On the other hand, \citet{2018arXiv181102588V} find that a 15\% deviation in planetary radius can just be originating from cooling effects of the planet itself.

This clearly implies that super-Earths can either be rocky or contain a significant fraction of water ice, but more analysis and observations are needed to clearly determine the fraction of rocky versus icy super Earths.

The formation of these super-Earth systems is subject of many studies \citep{2007ApJ...654.1110T, 2008ApJ...685..584I, 2009ApJ...699..824O, 2010ApJ...719..810I, 2010MNRAS.401.1691M, 2012ApJ...751..158H, 2014A&A...569A..56C,2014ApJ...780...53C, 2015A&A...578A..36O, 2016ApJ...817...90L, 2017MNRAS.470.1750I, Izidoro18}, where some studies additionally include tracking their composition \citep{2017MNRAS.464..428A, 2018MNRAS.479.3690B}. 

The simulations by \citet{2017MNRAS.470.1750I} combined the migration of already grown planets during the gas phase of the disc with long term evolution of the system after gas disc dispersal. During the gas-disc phase, the migrating planetary systems pile up in resonant chains anchored at the inner edge of the disc, which can then become unstable after the gas disc dissipates. These instabilities lead to small mutual inclinations between the surviving bodies, making the detection of the whole system via transit observations very hard. When mixing, for instance, 95\% of unstable system with 5\% of stable systems (which remain in resonance and coplanar), the Kepler dichotomy that mostly single planets are detected can be reproduced very nicely.

Expanding on this model, \citet{Izidoro18} included the accretion of the planetary cores via pebble accretion and found a similar trend that allows to reproduce fairly well the Kepler observations in terms of systems dynamical architecture and planet multiplicity. However, as in their model planetary cores form typically beyond the water ice line, the formed super-Earths are mostly of icy composition. Only if already interior rocky planets exist when icy super-Earths start to migrate inwards, is the formation of rocky close-in super-Earths possible. In this scenario, inward migrating icy super-Earths tend to encounter the rocky growing planets and shepherd them inwards \citep{2018MNRAS.479L..81R, Izidoro18}.

\citet{2008MNRAS.384..663R} proposed that a combination of planet composition (icy or rocky) and dynamical architecture could in principle allow to distinguish between different planet formation models. \citet{2016A&A...589A..15S} studied the water delivery to planetary embryos in the terrestrial planet region by icy pebble accretion, finding that the water ice content of the planets is greatly influenced by the inward movement of the water ice line.

It thus seems important to investigate where the first planetesimals form as this has great impact on the final planetary composition. Previous simulations seem to indicate that the water ice line might be the dominant location for the formation of the first planetesimals \citep{2013A&A...552A.137R, 2016ApJ...828L...2A, 2017A&A...608A..92D, 2017A&A...602A..21S}. Naively this would imply that all formed planets at this location should contain a large water ice content.

In this work here, we focus to calculate the water ice content of close-in super-Earth and how this allows to distinguish between the migration and in-situ formation scenario for hot super-Earths. For this we construct a simple model that takes planetary accretion and migration as well as disc evolution into account. In this work we do not model planet interiors so the word ``icy`` just refers to the water content of the planet, independent of what state of matter water would be.

Our work is structured as follows. In section~\ref{sec:methods} we explain our planet formation model and we show the results of our model in section~\ref{sec:nominal}, were we discuss different migration and ice line evolution models as well as different pebble accretion rates. We then discuss the implications of our findings in section~\ref{sec:disc} and summarize in section~\ref{sec:conclusions}.

\section{Methods}
\label{sec:methods}

In the following we present our model to study the composition of growing and migrating planets in evolving protoplanetary discs. For simplicity we assume that the planetary embryos only accrete solids and do not accrete a gaseous envelope, because hot super-Earths have difficulty retaining their atmosphere during the gas phase of the protoplanetary disc \citep{2017A&A...606A.146L, 2017MNRAS.471.4662C} and the very close-in hot super-Earths could lose their atmosphere through photoevaporation \citep{2014A&A...562A..80K, 2017ApJ...847...29O, 2018ApJ...853..163J}, giving rise to the discussion if super-Earths are rocky or icy in the first place. As our paper aims to understand what determines the bulk composition of the planetary core, we think not modeling gas accretion is justified.

\subsection{Disc model}

We use the disc model described in \citet{2015A&A...575A..28B}, which is based on 2D hydrodynamical simulations featuring viscous and stellar heating. This disc model is based on an $\dot{M}$ approximation, where the stellar accretion rate $\dot{M}$ decreases in time over 5 Myr using a fixed $\alpha$ parameter of $0.0054$. The decrease of $\dot{M}$ results in a reduction of gas surface density, which in turn reduces viscous heating and thus the disc's temperature. As a consequence the different ice lines move inwards in time, similar to previous works \citep{2007ApJ...654..606G, 2011ApJ...738..141O, 2015arXiv150303352B}. 

The evolution of the water ice line is initially very fast (top of Fig.~\ref{fig:NominalTemp}, where $t_0$ marks the initial time and $r_0$ the initial position), due to the fast diminishing of viscous heating, which moves the water ice line towards 1.2-1.3 AU in the first Myr \citep{2015A&A...575A..28B}. This evolution of the temperature will influence the chemical composition of the planet. Additionally, this disc model allows for zones of outward migration (see below), which are exterior to the water ice line \citep{2015A&A...575A..28B}. The same disc model was used in the planet formation simulations of \citet{2015A&A...582A.112B, 2016A&A...590A.101B, 2018MNRAS.474..886N} and in the N-body simulations of \citet{2017MNRAS.470.1750I, Izidoro18, Bitsch18}.

\subsection{Planetary growth}

Planetary embryos can grow very rapidly by accreting pebbles from the disc \citep{2010MNRAS.404..475J, 2010A&A...520A..43O, 2012A&A...544A..32L, 2012A&A...546A..18M}. The growth rates in these models then crucially depend on the amount of pebbles that are available in the disc \citep{2014A&A...572A.107L, 2018A&A...609C...2B}. Additionally, the sizes of the pebbles influence the accretion rates as well \citep{2012A&A...544A..32L, 2014A&A...572A.107L} in addition to the evolution of the pebble and gas surface density \citep{2018A&A...609C...2B}. 

Here, we instead use a simplified growth function that allows planetary embryos to grow from 0.01 Earth masses to 10 Earth masses within 1 Myr at 1 AU without planet migration. These growth rates are typical for pebble accretion \citep{Johansen2017}. We chose 0.01 Earth masses as starting mass of our planetary embryos, because pebble accretion takes over planetesimal accretion at this mass range \citep{Johansen2017}. The growth function is given by
 \begin{equation}
 \label{eq:growth}
 \dot{M}_{\rm core} = \left(\frac{M_{\rm core}}{6330 {\rm M}_{\rm E}}\right)^{2/3} \left(\frac{r}{1 {\rm AU}}\right)^{-0.5} \left(\frac{{\rm M}_{\rm E}}{\rm Myr}\right)\ ,
 \end{equation}
where the scaling with $M_{\rm core}^{2/3}$ corresponds to the 2D Hill accretion branch of pebble accretion \citep{2012A&A...544A..32L}. The radial power slope of $-0.5$ arises from the disc structure dependency \citep{Johansen2017}.

Water ice pebbles sublimate when drifting inwards of the water ice line. This reduces the amount of pebbles available to be accreted onto the planet and also changes the size of the pebbles. Recent simulations have assumed that the silicate pebbles that drift into the inner disc are only mm in size \citep{2015Icar..258..418M}, corresponding to the size of chondrules. This reduces the accretion rate. We assume here that the accretion rate onto planets is reduced by a factor of 4 if the planet is interior to the water ice line ($r<r_{\rm H_2O}$). We note that the growth rate of eq.~\ref{eq:growth} corresponds to the unreduced growth rate as if icy pebbles are available ($r>r_{\rm H_2O}$).

\begin{figure}
 \centering
 \includegraphics[scale=0.7]{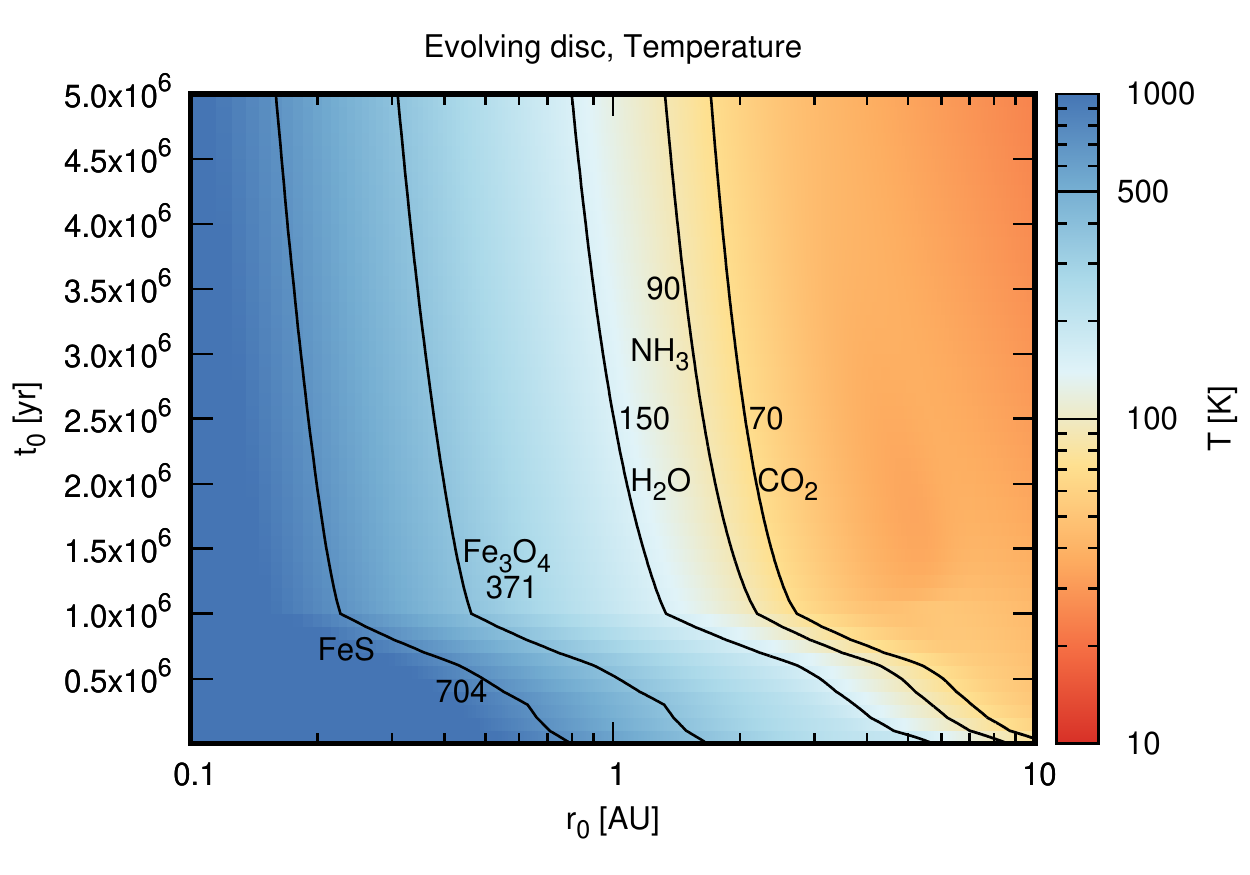}
 \includegraphics[scale=0.7]{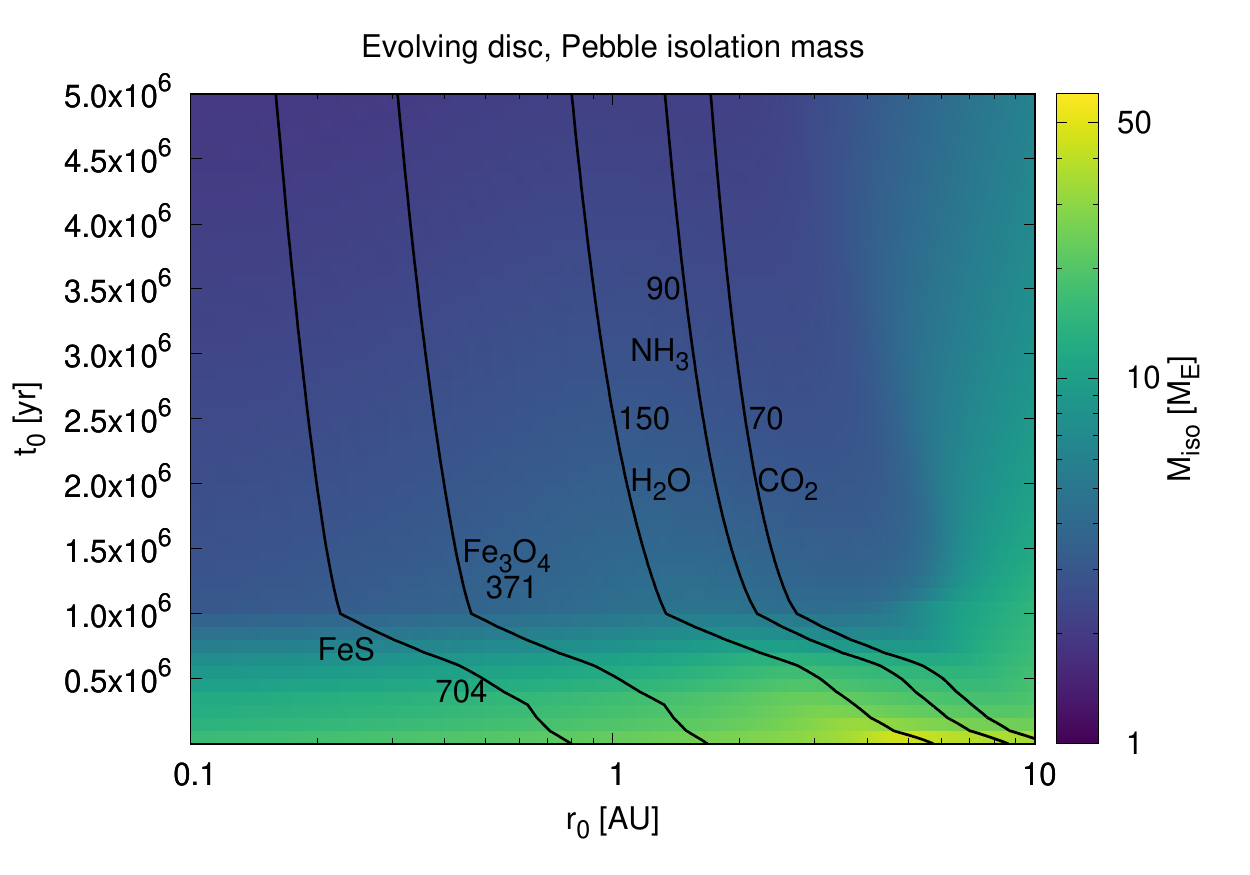}
 \caption{Temperature as a function of initial orbital distance $r_0$ and initial time $t_0$ (top) and the pebble isolation mass as a function of initial orbital distance $r_0$ and time $t_0$ (bottom) for the \citet{2015A&A...575A..28B} disc model. The black lines indicate the ice lines of FeS, Fe$_3$O$_4$, H$_2$O, NH$_3$ and CO$_2$, where the numbers mark their condensation temperatures, see Table~\ref{tab:species}. The temperature initially decreases very fast due to the decrease of viscous heating. This in turn reduces the pebble isolation mass in the inner disc. As the outer disc ($r>$6AU) is dominated by stellar irradiation, the aspect ratio and thus the pebble isolation mass remain large.
   \label{fig:NominalTemp}
   }
\end{figure}

Planet accretion stops at the pebble isolation mass, where the planets start to carve a small gap in the gas surface density, generating a pressure bump exterior to its orbit. In this pressure bump, pebbles can accumulate and do not reach the planet any more, which thus stops accreting pebbles. The pebble isolation mass depends on the disc's aspect ratio, viscosity and radial pressure gradient \citep{2014A&A...572A..35L, 2018arXiv180102341B}. We use here a simplified version of the pebble isolation mass which only depends on the disc's aspect ratio:
 \begin{equation}
  M_{\rm iso} = 25 \left(\frac{H/r}{0.05}\right)^{3} {\rm M}_{\rm E} \ .
 \end{equation}
We note that our disc model thus implies that the pebble isolation mass is large in the outer disc due to the larger aspect ratio. We show the pebble isolation mass as a function of radial position and initial time for our disc model in the bottom of Fig.~\ref{fig:NominalTemp}.

The disc structure leads normally to very small aspect ratios (around 2-2.5\%) in the inner disc (within 1 AU) at late times, which results in very small core masses of 1.6-3.0 Earth masses, which is a bit smaller than typical super-Earth planets (bottom of Fig.~\ref{fig:NominalTemp}). In this type of model, super Earths with larger masses can thus either form from collisions of small Earth mass bodies in the inner disc or by larger planets that migrate inwards from the outer disc. Bodies from the outer disc, however, have to cross the water ice line so their composition might be more water rich than bodies formed in the inner disc. In the following we investigate the interplay between accretion and migration speed as well as the evolution speed of the water ice line on the composition of the planet.

In our simulations, we only evolve single planets, so that effects of pebble filtering that could hinder the growth of interior planets \citep{2014A&A...572A.107L}, are not taken into account (see section~\ref{sec:disc}). Additionally, mutual interactions between planets can account for different migration behaviour (see section~\ref{sec:disc}).

\subsection{Planet migration}

Planets in gaseous discs interact gravitationally with the disc, exchanging angular momentum with the disc that leads to planetary migration (for a review see \citealt{2012ARA&A..50..211K} or \citealt{2013arXiv1312.4293B}). Here we include the effects of the Lindblad and corotation torques to calculate the planet migration rates. We follow the torque formula published by \citet{2011MNRAS.410..293P}, which gives good agreement to 3D hydrodynamical simulations \citep{2011A&A...536A..77B, 2015MNRAS.452.1717L}.

If the radial gradients in gas surface density and entropy, determined by the disc's temperature, are steep enough, planets can migrate outwards. This leads to zones of outward migration attached to the water ice line in our disc model \citep{2015A&A...575A..28B}. In the here presented work we will additionally test two other migration scenarios, (i) an in-situ planet formation scenario, where the planets do not migrate and their composition is strictly given by their position relative to the ice lines and (ii) where planets are only allowed to migrate inwards, due to a low viscosity\footnote{This assumption is based on new results of disc evolution \citep{2016ApJ...821...80B, 2016arXiv160900437S}, where the angular moment is transported via disc winds, while the midplane remains laminar with low viscosity, resulting in inward migration of planets \citep{2018arXiv180511101K, 2018ApJ...864...77I}.}. Low viscosity will saturate the corotation torques, even when large radial gradients in entropy are present \citep{2008ApJ...672.1054B} and planets will still migrate only inwards.

However, if multiple planets are present in the protoplanetary discs, they excite their eccentricities which can quench the entropy driven corotation torque \citep{2010A&A.523...A30} and stop outward migration. A chain of small mass planets, where single planets would migrate outwards, would then migrate inwards \citep{2013A&A...553L...2C, 2017MNRAS.470.1750I, Izidoro18}. This could have important consequences for the composition of the individual planets, but is not modeled here and will be investigated in future work.

\subsection{Chemical composition of planets}

In order to account for the chemical composition of the planet, we include only the major rock and ice forming species. The mixing ratios (by number) of the different species as a function of the elemental number ratios is denoted X/H and corresponds to the abundance of element X compared to hydrogen for solar abundances, which we take from \citet{2009ARA&A..47..481A} and are given as follows: He/H = 0.085; C/H = $2.7\times 10^{-4}$; N/H = $7.1\times 10^{-5}$; O/H = $4.9\times 10^{-4}$; Mg/H = $4.0\times 10^{-5}$; Si/H = $3.2\times 10^{-5}$; S/H = $1.3\times 10^{-5}$; Fe/H = $3.2\times 10^{-5}$.

These different elements can combine to different molecular species. We list these species, as well as their condensation temperature and their volume mixing ratios $v_{\rm Y}$ in Table~\ref{tab:species}. More details on the chemical model can be found in \citet{2017MNRAS.469.4102M} and \citet{2018MNRAS.479.3690B}.

We note that we only change the accretion rate of a planet at the water ice line. For all other ice lines, the accretion rate is unaffected. This then also implies that as the total mass a planet accretes in each timestep is the same, the composition of the accreted material solely depends on the planets position relative to the ice lines. The amount of each molecule that is accreted then follows their relative abundances to each other as stated in Table~\ref{tab:species}. This also means that planets formed in a disc temperature range of 90-150K accrete the largest water ice fraction by mass in our model, corresponding to $\sim$35\%. The exact planetary composition is related to our chemical model, where the other parts of the planetary composition by mass for planets just formed at $r>r_{\rm H_2O}$ correspond to Fe$_3$O$_4$ (7.5\%), FeS (12.3\%), Mg$_2$SiO$_4$ (11.9\%), Fe$_2$O$_3$ (7.8\%) and MgSiO$_3$ (25.5\%). Planets formed at T<90K additionally accrete NH$_3$, reducing the fraction of all other elements inside each pebble and thus also of the planet they accrete on. This means that planets forming at T<90K will have a smaller abundance of each molecule compared to planets forming just at T>91K, but will feature NH$_3$.

Super-Earths are found around host stars of different types with various metallicities and element abundances \citep{2012Natur.486..375B, 2014Natur.509..593B, 2018ApJ...867L...3B}. Differences in the chemical abundances, especially of oxygen, can lead to different formation pathways of super-Earths \citep{2016A&A...590A.101B}, but we will keep solar abundances for all chemical elements for this work. Our model is also designed for solar type stars, but could be easily expanded for different stellar types. We note that the maximum water ice fraction by mass planets can have in our model is $\sim$35\%, which is actually not too far away from the proposed 20\% water content for super-Earths by \citet{2018arXiv181103202G}. This value could also change for exoplanet systems due to abundance differences for the different chemical elements, making detailed stellar abundance studies of planet host stars very important for planet formation studies.

\begin{table*}
\centering
\begin{tabular}{c|c|c}
\hline
Species (Y) & $T_{\text{cond}}$ {[}K{]} & $v_{\text{Y}}$ \\ \hline \hline
CO & 20  & 0.45 $\times$ C/H (0.9 $\times$ C/H for $T < 70$ K) \\[5pt]
CH$_4$ & 30 & 0.45 $\times$ C/H (0 for $T > 70$ K)  \\[5pt]
CO$_2$ & 70 & 0.1 $\times$ C/H  \\[5pt]
NH$_3$ & 90* & N/H  \\[5pt]
H$_2$O & 150 & O/H - (3 $\times$ MgSiO$_3$/H + 4 $\times$ Mg$_2$SiO$_4$/H + CO/H \\
& & + 2 $\times$ CO$_2$/H + 3 $\times$ Fe$_2$O$_3$/H + 4 $\times$ Fe$_3$O$_4$/H) \\[5pt]
Fe$_3$O$_4$ & 371 & (1/6) $\times$ (Fe/H - S/H) \\[5pt]
C (carbon grains) & 500 & 0 \\[5pt]
FeS & 704 & S/H \\[5pt]
Mg$_2$SiO$_4$ & 1354 & 0.75 $\times$ Si/H  \\ [5pt]
Fe$_2$O$_3$ & 1357** & 0.25 $\times$ (Fe/H - S/H) \\ [5pt]
MgSiO$_3$ & 1500 & 0.25 $\times$ Si/H  \\  \hline
\end{tabular}
\caption[Condensation temperatures]{Condensation temperatures and volume mixing ratios of the chemical species. Condensation temperatures for molecules from \citet{2003ApJ...591.1220L}. *Condensation temperature for NH$_3$ from \citet{2015A&A...574A.138T}. For Fe$_2$O$_3$ the condensation temperature for pure iron is adopted \citep{2003ApJ...591.1220L}. Volume mixing ratios $v_{\rm Y}$ (i.e. by number) adopted for the species as a function of disc elemental abundances (see e.g. \citealt{2014ApJ...791L...9M}).}
\label{tab:species}
\end{table*}

\section{Composition of planets}
\label{sec:nominal}

Even though our model allows to study the detailed composition of planets, we will focus here only on the water ice fraction that the planet has accreted. We are especially interested in the interplay between migration and the ice line evolution, so we present in the following the different scenarios of the interplay between migration and ice line evolution. We then also discuss scenarios with only inward migration and different accretion rates.

\subsection{No ice line evolution without planet migration}

In this scenario the water ice line is fixed in time and the planets do not migrate. The planetary composition is thus solely determined by the planetary starting position and if the planetary embryo is initially placed inside or outside the water ice line.

\begin{figure}
 \centering
 \includegraphics[scale=0.7]{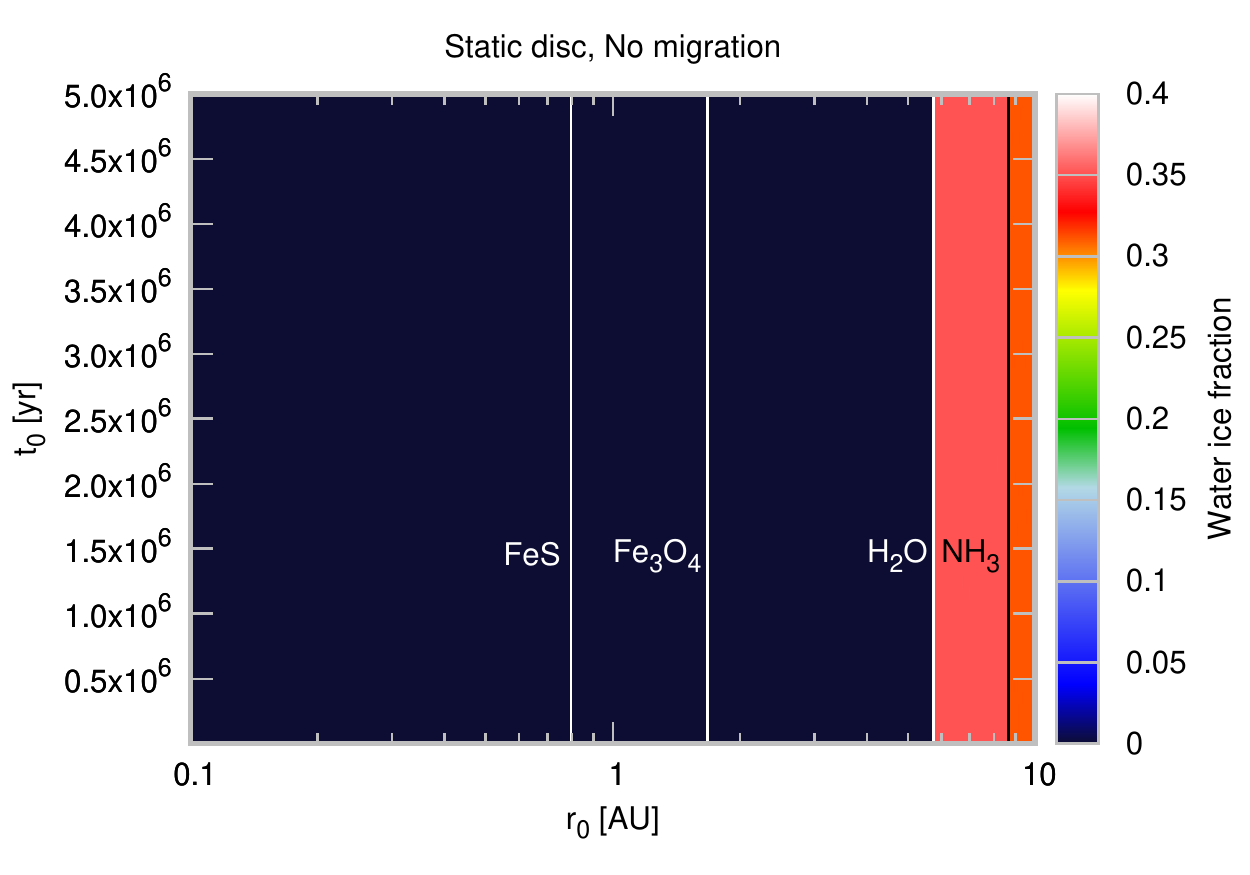}
 \caption{Water ice fraction of the planets for a model where the water ice line is fixed at $\sim$5.5 AU and the planets do not migrate. The labels of $r_0$ and $t_0$ indicate where and when a planetary embryo was introduced into the disc. The white and black lines correspond to the lines of FeS, Fe$_3$O$_4$, H$_2$O, and NH$_3$. Only planets forming exterior to $\sim$5.5 AU (at T<150K) can contain water ice. These planets contain by mass about $\sim 35$\% of water ice, which is the maximum allowed in our chemical model.
   \label{fig:icenomignodisc}
   }
\end{figure}

In Fig.~\ref{fig:icenomignodisc} we present the water ice fraction (ice to rock ratio) of the planets formed in our model, where the water ice line is fixed at $\sim$5.5 AU and the planets do not migrate. Obviously only planets in the parts of the disc where T<150 K contain water ice. These planets actually contain $\sim 35$\% of water ice, which is the maximum allowed in our model. Planets formed interior to the water ice line then contain, by definition of our model, no water ice at all and consist of a rocky composition. Planets formed exterior to the NH$_3$ ice line at 90K contain less water ice, because now additionally NH$_3$ ice is accreted reducing the amount of water ice accreted.

This already indicates that in the situation of a non evolving disc and no planet migration that the chemical composition of a formed planet is solely dependent on the underlying chemical model. Additionally, it indicates that in this formation scenario super-Earths with a presumable icy component like GJ1214b \citep{2013ApJ...765..127F} or HAT-P-11b \citep{2014Natur.513..526F} could only be formed if they were scattered to the inner disc. These two planets have eccentricities of around $\sim$0.2, indicating that dynamical instabilities and scattering events might actually have occured in these systems. Although some super-Earths may have been scattered to the inner disk by gas giant planets \citep{2011A&A...530A..62R}, it is very unlikely this scenario can account for the majority of super-Earths systems.  Super-Earths systems have low orbital eccentricities ($\sim$0.1) and low
 inclinations of up to a few degrees \citep{2011arXiv1109.2497M, 2011ApJS..197....1M, 2016PNAS..11311431X, 2018arXiv180700549V}, indicating that high eccentricity/inclination scattering events by outer gas giant planets
 \citep{2011A&A...530A..62R} are probably very rare in these systems and can potentially not account for the majority of water rich super-Earths. Super-Earths orbital eccentricities and inclinations are more likely  outcomes of dynamical instabilities and scattering events among low mass planets \citep{2017MNRAS.470.1750I, Izidoro18}.

\subsection{No ice line evolution with inward/outward planet migration}

\begin{figure}
 \centering
 \includegraphics[scale=0.7]{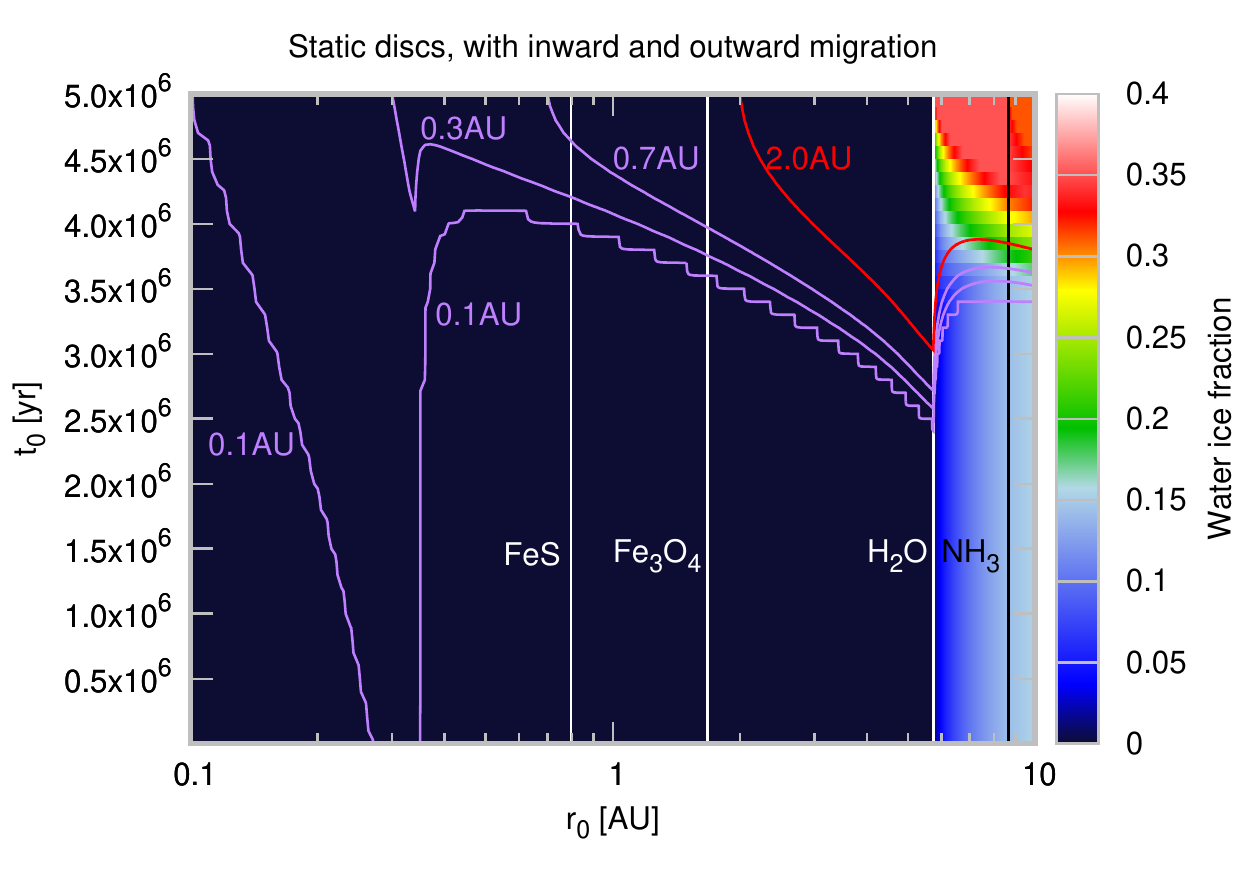}
 \caption{Same as Fig.~\ref{fig:icenomignodisc}, but planets are allowed to migrate. The purple and red lines mark the final positions of the planets. Planets in between the lines marked with 0.1 AU and 0.3 AU have a final position between 0.1 and 0.3 AU. Most planets migrate into the inner edge of the disc at 0.1 AU. Planets forming exterior to the water ice line ($r>r_{\rm H_2O}$) first accrete water rich material, but then migrate inwards accreting mostly rocky material resulting in a low water ice fraction of the planet. This results in a composition gradient of the planets with respect to the water ice line.
   \label{fig:icemignodisc}
   }
\end{figure}

In Fig.~\ref{fig:icemignodisc} we present the water ice fraction of the planets, where the water ice line is fixed at $\sim$5.5AU meaning that we do not evolve the disc at all, but the planets are allowed to migrate. Planets that originate from the outer disc accrete first water rich material, but eventually migrate inwards across the water ice line, where they then only accrete rocky material\footnote{Our disc model allows outward migration, but as we keep the disc structure fixed at t=0, planets in the outer disc need to be around 20${\rm M}_{\rm E}$ to migrate outwards \citep{2015A&A...575A..28B}. However, before the planets can reach these masses they have already migrated to the inner disc.}. This leads to a water fraction that can be just a few percent, even for planetary embryos originating in the water ice rich parts of the disc. Planets forming at late times are staying relatively small, so that their migration is negligible, and they never cross the water ice line. These planets thus accrete the maximum water fraction of $\sim 35$\% allowed in our model.

Inward migration of planets seems to be key in order to form planets with different water content (alongside disc evolution, see below) within the same system. Our simulations here predict that the water ice content should be well below $\sim$35\%, unless the planets grow to pebble isolation mass before migrating across the water ice line. This happens either for late formation times or if planets form far out in the protoplanetary disc.

The mechanism of forming planets at the water ice line and then migrating them inwards was also proposed by \citet{2017A&A...604A...1O} to explain the formation of the Trappist-1 system \citep{2016Natur.533..221G, 2017Natur.542..456G}. A formation scenario like this would imply that planets formed directly at the water ice line have a similar composition and thus, if no atmosphere is present, a similar density. The densities of the planets in the Trappist-1 system are roughly 0.6-1.0 $\rho_{\rm E}$ \citep{2017Natur.542..456G}, indicating that a larger fraction of water compared to Earth could be present in the planets supporting this scenario. 

The planets Trappist-1c and Trappist-1e seem to have a density similar to Earth, indicating a rocky nature of the planet, while Trappist-1b, d, f, g and h require envelopes of volatiles in form of thick atmospheres, oceans or ice, with a water mass fraction of less than 5-10\% \citep{2018A&A...613A..68G}. This could imply that the planets for Trappist-1 formed at different distances to the water ice line and thus Trappist-1c and 1e probably very close to it and always stayed interior to the water ice line.

\subsection{Ice line evolution without planet migration}

\begin{figure}
 \centering
 \includegraphics[scale=0.7]{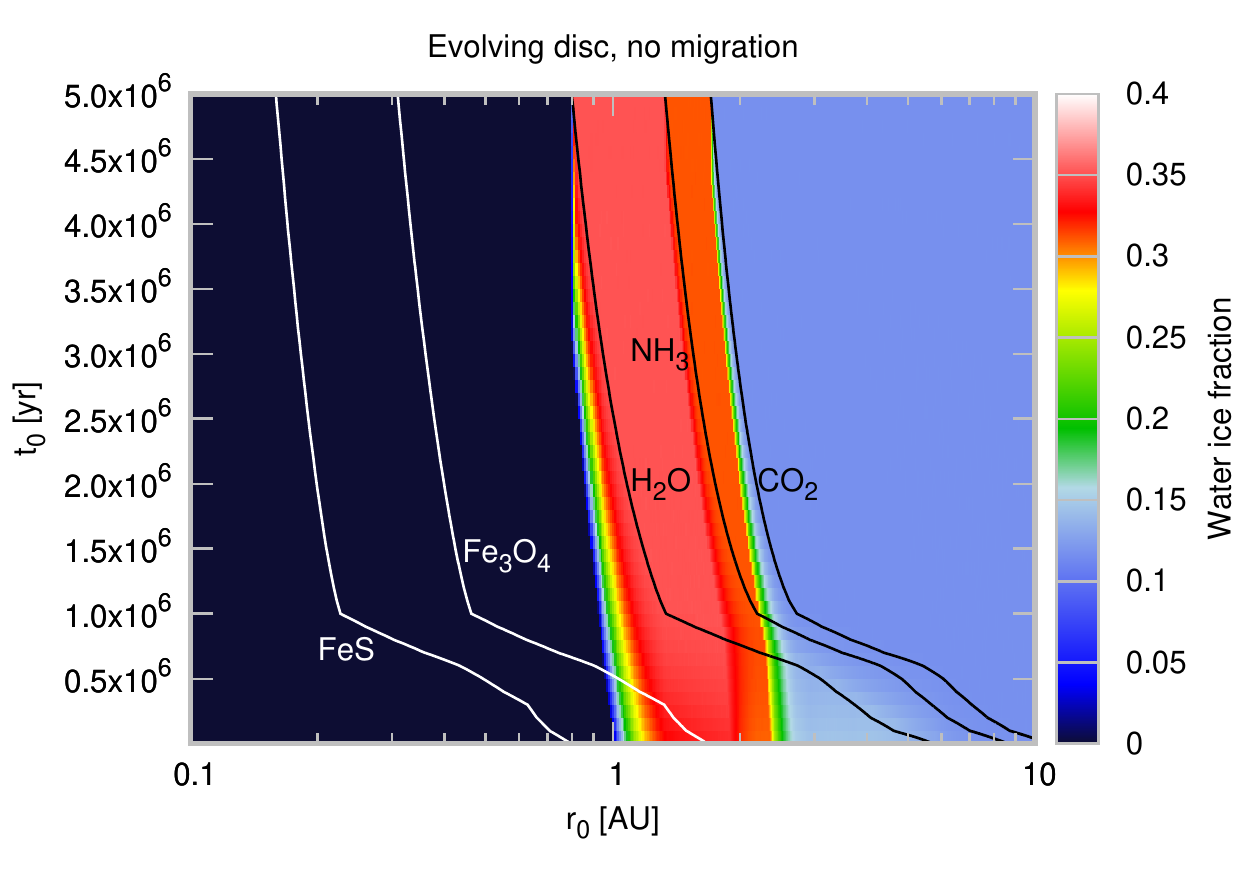}
 \caption{Water ice fraction of the planets for a model where the water ice line evolves in time as indicated in Fig.~\ref{fig:NominalTemp} and the planets are not migrating. The black and white lines correspond to the temperature at the given time $t_0$. As the growth of the planets takes several Myr, the ice line sweeps over the planet's location also if the planet is initially at $r<r_{\rm H_2O}$. Only planets forming exterior to T=150K can contain water ice. These planets can contain up to about $\sim$35\% of water ice, which is the maximum allowed in our chemical model. In the outer part at T<70K, the water ice fraction decreases again, because our chemical model allows for more CO production which binds oxygen efficiently resulting in a smaller water ice fraction.
   \label{fig:icenomigdisc}
   }
\end{figure}

In Fig.~\ref{fig:icenomigdisc} we show the water ice fraction of planets where the water ice line evolves in time, but the planets are fixed to their initial orbit and do not migrate. Only planets that are situated at $r>r_{\rm H_2O}$ or are swept by the water ice lines' inward movement can contain water. Nevertheless, we observe a large diversity in the water ice fractions for our planets. Planets that are swept by the water ice line only at the end of their growth phase contain a water fraction less than $\sim$35\%. Only planets that are already at $r>r_{\rm H_2O}$ or close to $r_{\rm H_2O}$ contain these large water fractions. Surprisingly, most planets in our model either only contain a large water fraction or no water at all. 

Planets at larger orbital distances actually contain less than $\sim$35\% water ice, even though water ice is always in solid form. This is related to two different ice-lines that also move inwards in time: the NH$_3$ and the CO$_2$ ice line. As the NH$_3$ ice line sweeps over the planets, they start to accrete NH$_3$ ice. This process reduces the water ice fraction, because the planet accretes more different materials, but still accretes at the same growth rate. We note that we do not include any effects of different growth rates for the CO$_2$ or NH$_3$ ice lines, only for the water ice line.

When the CO$_2$ ice line sweeps over the planets the same process happens, namely that planets accrete now additionally CO$_2$ ice, reducing the water ice fraction. Additionally our chemical model increases the amount of CO when T<70K. This excess CO reduces the amount of water ice, due to the binding of oxygen in CO leaving less oxygen for the water ice. This greatly reduces the water ice abundance in the planet (light blue in Fig.~\ref{fig:icenomigdisc}).

Our simulations also show that planets that are originally interior to the water ice line can accrete water ice due to the inward movement of the water ice line, similar as in \citet{2016A&A...589A..15S}. Planetary embryos that are formed at an orbital distance down to the final position of the water ice line at 5 Myr will accrete water in our model. The water ice fraction then depends slightly on how fast the planets can grow in respect to the inward movement of the water ice line. This corresponds to the planets that are interior to the 150K line marked in Fig.~\ref{fig:icenomigdisc}. Similar effects happen for the other ice lines.

\subsection{Ice line evolution with inward/outward planet migration}

\begin{figure}
 \centering
 \includegraphics[scale=0.7]{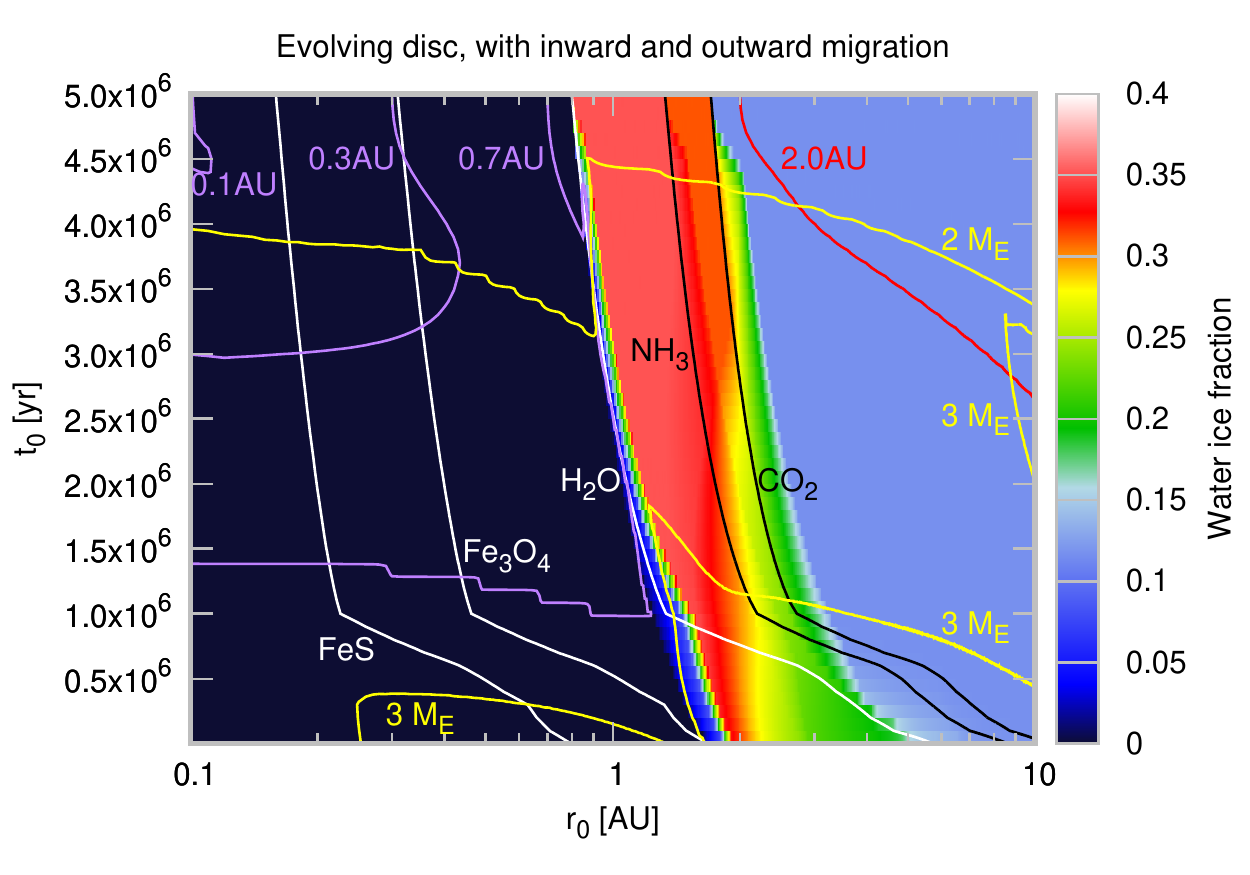}
 \caption{Same as Fig.~\ref{fig:icenomigdisc}, but planets are allowed to migrate through the disc. The yellow lines mark the final masses of the planets which around 2-3 Earth masses. Due to our disc model, planets exterior to the water ice line ($r>r_{\rm H_2O}$) migrate outwards and stay exterior to the water ice line until disc dissipation. This thus results in very similar water ice fractions in the planets compared to Fig.~\ref{fig:icenomigdisc} where planets do not migrate. Only planets that form close to the central star at late times have orbits interior of 0.7 AU. Planets forming earlier and farther out are caught in the region of outward migration, which leaves them stranded at exterior orbits in the region of outward migration (yellow line in Fig.~\ref{fig:migmap}). 
   \label{fig:icemigdisc}
   }
\end{figure}

In Fig.~\ref{fig:icemigdisc} the water ice line evolves in time, but also the planets are allowed to migrate. Here the water ice fraction in the cold parts of the disc (T<90K) evolves due to the process mentioned above. As the planets start to grow, they migrate through the disc. 

In the used disc model of \citet{2015A&A...575A..28B}, planets formed exterior to the water ice line can migrate outwards, if they become large enough (see the yellow growth track in Fig.~\ref{fig:migmap}). In contrast to Fig.~\ref{fig:icemignodisc} the disc evolves in time and thus the region of outward migration shrinks in time and also smaller planets can experience outward migration \citep{2015A&A...575A..28B}. In this disc model, the pebble isolation mass, which is the maximum mass planets can grow to in our model, results in planetary masses that can experience outward migration. Thus planets that start growing by pebble accretion once the water ice line has swept over them, will stay in the cold part of the disc until disc dissipation and thus they all have a large water ice content.

It seems that the water ice fraction of the planets in the case of outward migration is very similar to the case when the planets do not migrate at all (Fig.~\ref{fig:icenomigdisc}). However, the final planetary masses are slightly lower in the migration case, because planets always migrate to the minimum in H/r \citep{2013A&A...549A.124B}, which is also the minimum of the pebble isolation mass and hence planetary mass.

Hot super-Earths with large ice content could thus only be formed by either scattering from the outer disk into the inner parts or by chains of planets that migrate inwards together\footnote{Mutual interactions of multiple bodies increase their eccentricities, which reduces the contribution of the entropy related corotation torque \citep{2010A&A.523...A30}, which can allow the chain of planets to migrate inwards together \citep{2013A&A...553L...2C}.}. We note that even rocky planets that form early (<1.5 Myr) in the inner disc end up at orbits exterior to 0.7 AU (see green line in Fig.~\ref{fig:migmap}). This is related to the evolution of the region of outward migration in the disc, which allows planets of around 2-5 Earth masses to migrate outwards at the late stages of the disc evolution, even when they are in the inner disc \citep{2015A&A...575A..28B}, as invoked in \citet{2016MNRAS.458.2962R} to explain the formation of Jupiter's core. This leads to only a small fraction of parameter space that allows the formation of close-in rocky super-Earths. Multi-body dynamics, on the other hand, can change the late migration history of the growing planets and is discussed in section~\ref{sec:disc}.

\subsection{Ice line evolution with only inward migration}

\begin{figure}
 \centering
 \includegraphics[scale=0.7]{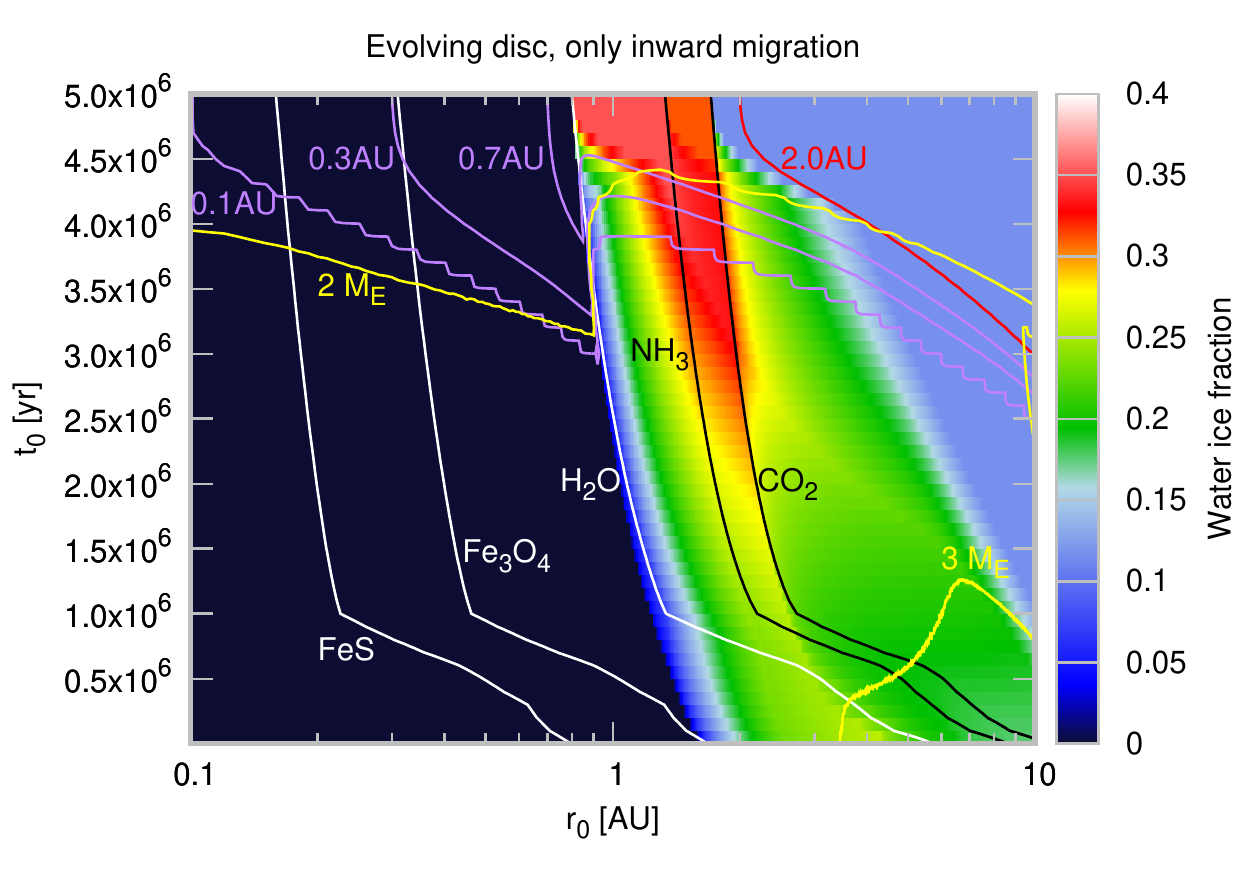}
 \caption{Same as Fig.~\ref{fig:icemigdisc}, but planets only migrate inwards. Planets can then start to grow exterior to the water ice line ($r>r_{\rm H_2O}$) and then migrate inwards resulting in a water ice gradient of the planets depending on their exact formation position in respect to the water ice line. Planets forming before 3-3.5 Myr all end up migrating to the inner edge at 0.1 AU (red/purple lines) in contrast to Fig.~\ref{fig:icemigdisc}, because planet migration is only directed inwards. The final planetary masses though, are very similar to the previous simulations.
   \label{fig:icelowmigdisc}
   }
\end{figure}

In this section, we only allow inward migration of planets and suppress outward migration, but keep the disc evolution the same as in the previous section and present the water ice fraction in Fig.~\ref{fig:icelowmigdisc}. The migration rates are calculated using $\alpha_{\rm mig}=0.0001$, which describes the viscosity used for planet migration. This value is low enough to prevent outward migration through the entropy driven corotation torque. The planetary embryos migrate inwards faster than the water ice line evolves. This can be seen by two effects:
\begin{itemize}
 \item[1)] The planetary embryos starting interior to the water ice line (interior to the 150K line) never accrete water, except in the early stages when the water ice line evolves inward faster than the planets migrate. Even if the water ice line sweeps eventually over the initial planetary starting position later on, the planet will not accrete water ice, because it already migrated inwards further, because the inward migration is faster than the water ice line evolution. This happens, for example, for the planetary embryo starting at 1 AU at $t_0$=1Myr.
 \item[2)] Planetary embryos forming in the cold part of the disc (T<150K) close to the water ice line do not accrete the maximum fraction of water ice ($\sim$35\%). Instead as they start to grow, their migration speed increases and they can migrate inwards of the water ice line where they then finish their formation by accreting rocky materials, thus reducing their water ice component. 
\end{itemize}
This clearly illustrates that the evolution of the water ice line and the migration of planets play in unison an important role in determining the water ice fraction of formed super-Earths. Planetary systems with multiple super-Earths, where planets have different densities are thus a signpost of the inward movement of the water ice line or of planet migration (or both).

Additionally, it clearly shows that if regions of outward migration are attached to the water ice line and if planet formation starts at the water ice line, the born planets should have a large water ice content. Smaller water ice contents are only possible if planets form at or beyond the water ice line, but then migrate inwards during their growth eventually entering the hotter regions of the disk (T>150K).

\subsection{Faster and slower growth rates}

\begin{figure}
 \centering
 \includegraphics[scale=0.7]{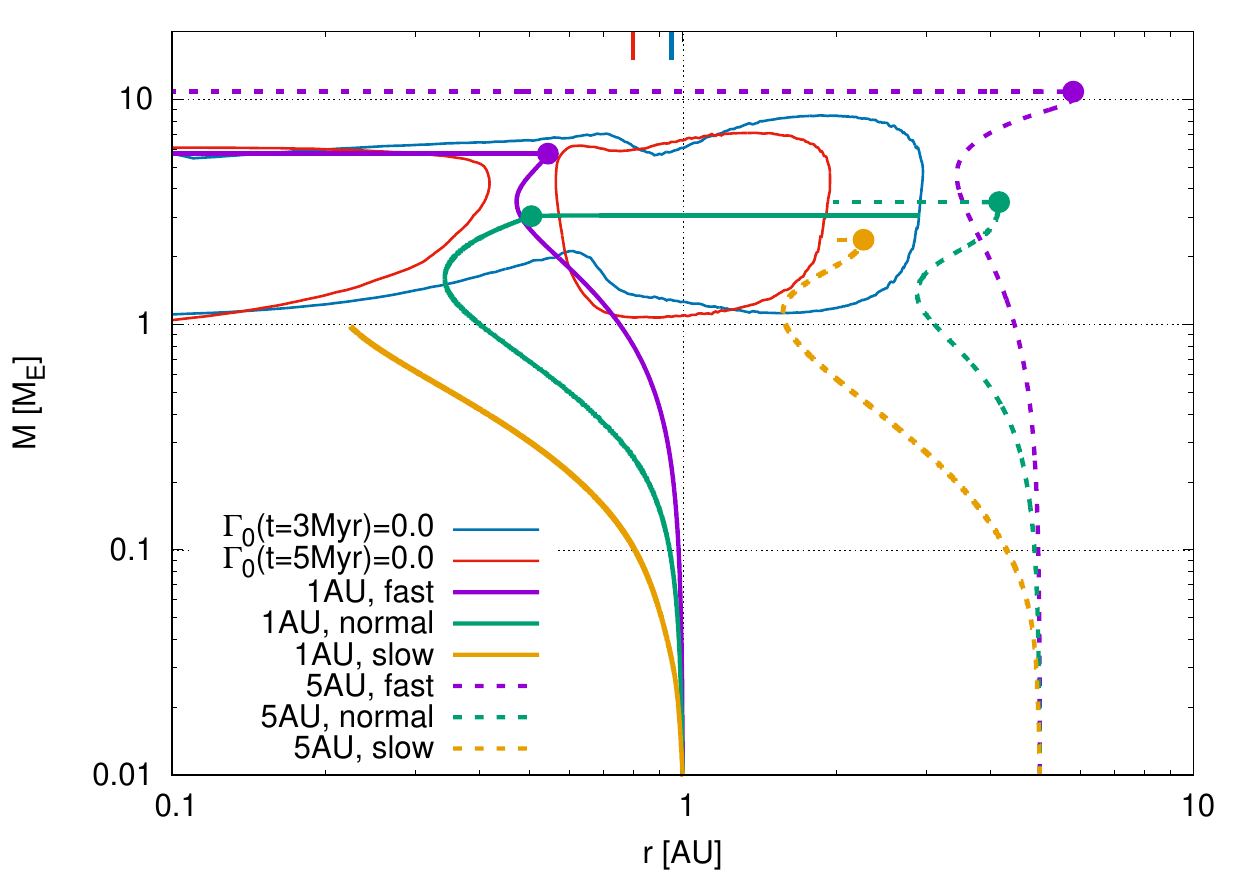}
 \caption{ Growth tracks of six planetary embryos, which start at either 1 or 5 AU, where three different growth scenarios are invoked: the normal one (eq.~\ref{eq:growth}) and a growth that is 4 times faster and 4 times slower. The planets start at t=0 Myr. The dots mark the pebble isolation mass, when planets stop growing. The planet growing at 1 AU with slow growth never reaches pebble isolation mass until the end of the discs lifetime. The solid contours mark the regions of outward migration at disc ages of 3 and 5 Myr, where planets inside of these contours can migrate outwards. All growth tracks are for scenarios where planets are allowed to migrate outwards. The small tics on the top mark the position of the water ice line at 3 and 5 Myr. The green growth track (1AU, normal growth) features an interesting trajectory: the planet first migrates all the way out to 3 AU at 3 Myr, but then follows the zero torque region ($\Gamma_0$) and migrates inwards with it. The planet is thus rocky in composition (pebble isolation reached interior of 1 AU), but its final position is exterior to the water ice line ($r>r_{\rm H_2O}$).
   \label{fig:migmap}
   }
\end{figure}

\begin{figure}
 \centering
 \includegraphics[scale=0.7]{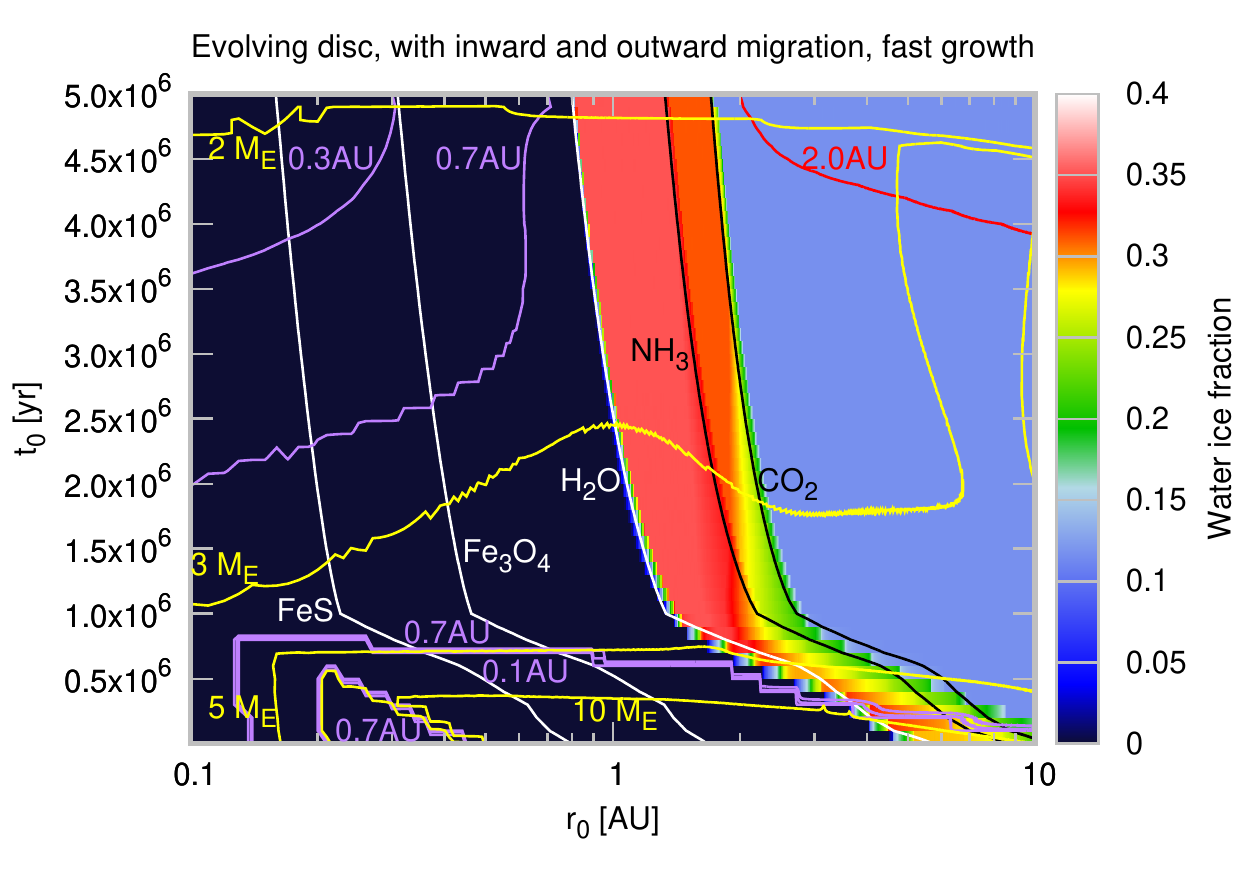}
 \includegraphics[scale=0.7]{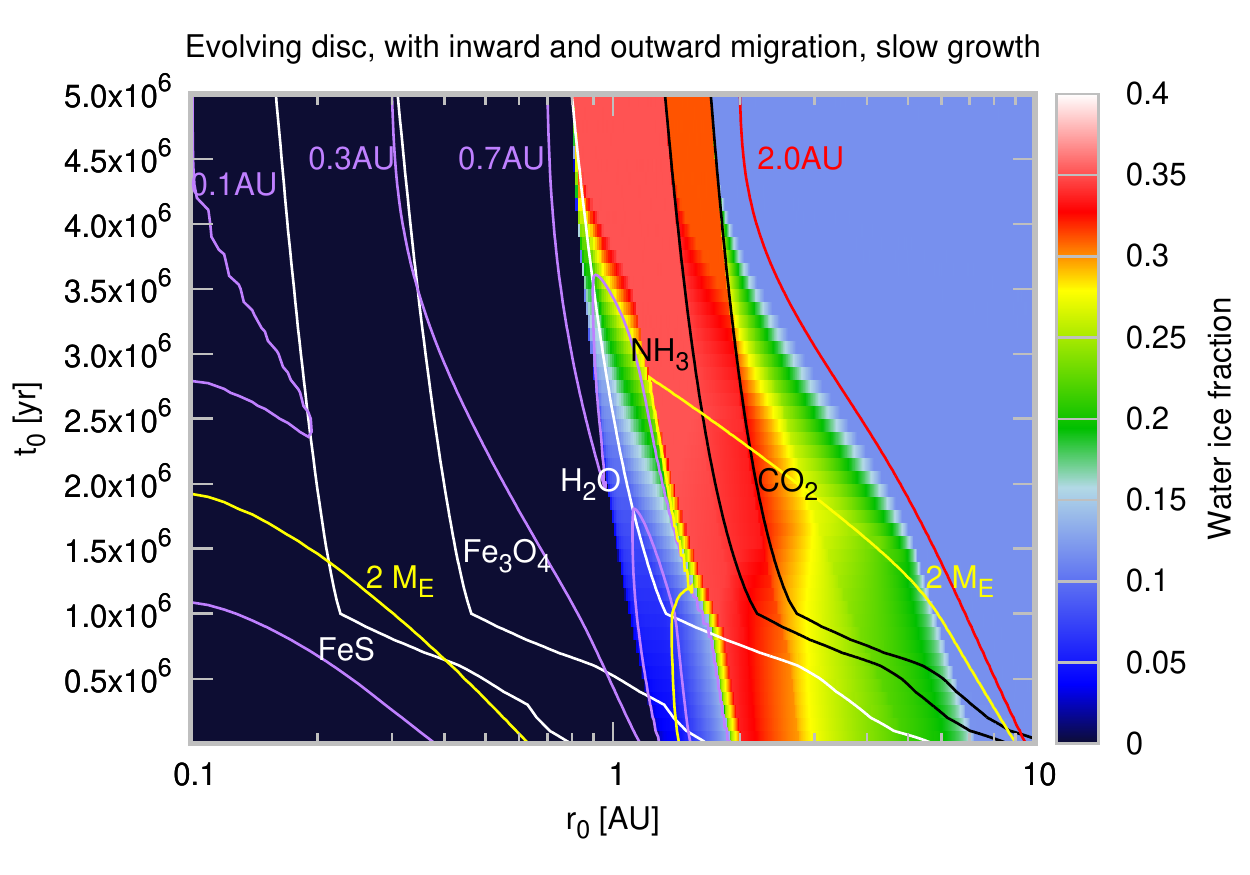}
 \caption{ Same as Fig.~\ref{fig:icemigdisc}, but planets grow with a 4 times faster growth rate (top) or 4 times slower growth rate (bottom). For faster growth rates planets grow bigger, especially in the early stages of the disc evolution when the pebble isolation mass is still large. Additionally the fast growth also interior to the water ice line allows planets to reach large masses and migrate inwards even before they are swept by the water ice line. Slow growth shows exactly the opposite behaviour, namely that planets stay small and that they will not migrate significantly before they are swept by the water ice line.
   \label{fig:fastslow}
   }
\end{figure}

The results of the previous simulations imply that migration and ice line evolution play a crucial role in determining the water ice content of forming planets. Additionally, the accretion rate is important, because it determines how much a planet that forms interior to the water ice line at $r<r_{\rm H_2O}$ can grow before it is eventually swept by the water ice line.  In Fig.~\ref{fig:migmap} we show the growth tracks of planets featuring different growth rates and in Fig.~\ref{fig:fastslow} we show the water ice content of growing planets formed with 4 times faster (top) and 4 times slower (bottom) growth rates.

Fast accretion allows planets that form interior, but close to the water ice line, to grow fast enough to migrate inwards before their location is swept by the water ice line. Therefore a clear distinction in the water ice content related to the water ice line is visible. 

In Fig.~\ref{fig:icemigdisc} planetary embryos that are introduced before 1.5 Myr in the inner regions of the disc grow to become rocky planets with orbits exterior to 0.7 AU. The growth track of such a planet is shown in Fig.~\ref{fig:migmap} marked by the solid green curve. If the planets, on the other hand, grow faster, they reach a higher pebble isolation mass at an early disc evolution stage, and attain a mass that is too high for outward migration. The planets thus migrate all the way to the inner edge of the disc (see the solid purple growth track in Fig.~\ref{fig:migmap}) and stay there until disc dissipation. Fast growth exterior to the water ice line allows the planets to grow bigger and they would also migrate all the way to the inner disc, because they can not be contained by the region of outward migration at the late stages of the disc evolution (dashed purple line in Fig.~\ref{fig:migmap}).

Slower planetary accretion produces more planets with non-zero water ice contents, because the ice line's fast evolution sweeps across a large fraction of the disc before planets can migrate away. On the other hand, planets that are initially located at $r>r_{\rm H_2O}$ will have a lower water ice content if they only migrate inwards (Fig.~\ref{fig:icelowmigdisc}), because they migrate inwards earlier into the hot inner disc, resulting in the accretion of more rocky material like in Fig.~\ref{fig:icemigdisc}. Additionally, the planets growing slowly close to the water ice line (bottom in Fig.~\ref{fig:fastslow}) never migrate all the way to the inner disc edge, as also in the nominal growth case (Fig.~\ref{fig:icemigdisc}). In fact in some cases the growth is so slow that planets in the inner disc do not even reach pebble isolation mass (see the growth track marked in solid yellow in Fig.~\ref{fig:migmap}).

We thus conclude that the pebble accretion rate can have an important influence on the water ice content of the planets, but as long as outward migration is allowed, this will mostly influence planets exterior to a final position of 0.7 AU, where it is very hard to probe the water ice composition of planets via observations. Rocky planets that end up at the inner disc edge have to form in this scenario interior to the water ice line and they have to form early with high pebble accretion rates, as all other mechanisms do not allow rocky planets to migrate into the inner edge. In the case of fast accretion rates (top in Fig.~\ref{fig:fastslow}) some planets (starting at $r_0>$5AU and $t_0<$300kyr) with large water ice contents migrate towards the inner disc edge. However these planets are quite massive (around 10 Earth masses), implying that they should undergo gas accretion in some form. These planets are then too massive to have lost their atmosphere via the interactions with their host star, making a composition determination not possible.

This result might of course change when taking multi-body dynamics into account, because eccentricity excitations quench the corotation torque preventing outward migration \citep{2010A&A.523...A30}. On the other hand, if only inward migration is allowed (Fig.~\ref{fig:icelowmigdisc}) purely rocky planets and planets with different water ice contents can easily migrate towards the inner disc edge.

We note that the N-body simulations of \citet{Izidoro18} actually require for the formation of rocky super-Earths a fast enough growth for embryos initially located at $r<r_{\rm H_2O}$. The requirement of fast early growth in \citet{Izidoro18} is due to the outward migration of planets that are initially located at $r<r_{\rm H_2O}$ and are then swept by the ice line (planets below the H$_2$O line at t$<$1 Myr in Fig.~\ref{fig:icemigdisc}) and would then accrete water rich material.

\section{Discussion}
\label{sec:disc}

In this section we discuss first some caveats and then some implications of our model on the composition of super-Earths planets. We show in Table~\ref{tab:Results} the summary of our simulations and their results.

 {%
 \begin{table*}
  \centering
  \begin{tabular}{cccc}
  \hline \hline
  Ice line evolution & Planet migration & Water content of inner planets (r<0.3 AU) & Figure  \\ \hline
  No & No & Only rocky by construction, no water rich planets & Fig.~\ref{fig:icenomignodisc}  \\
  No & Yes & Mostly rocky, some water rich planets originating from beyond $r_{\rm H_2O}$ & Fig.~\ref{fig:icemignodisc}  \\
  Yes & No & Only rocky by construction, no water rich planets & Fig.~\ref{fig:icenomigdisc}   \\
  Yes & Yes & Only rocky planets, water rich planets trapped outside of 0.7 AU & Fig.~\ref{fig:icemigdisc} \\
  Yes & only inwards & Rocky and icy planets with composition gradient depending on $r_{\rm 0}$ & Fig.~\ref{fig:icelowmigdisc}  \\
  Yes & Yes & fast growth allows rocky interior planets, no inner water rich planets & Fig.~\ref{fig:fastslow} (top)  \\
  Yes & Yes & slow growth allows rocky interior planets, no inner water rich planets & Fig.~\ref{fig:fastslow} (bottom)  \\  
  \hline
  \end{tabular}
  \caption{ Summary of our simulation results for the different models. Observational predictions imply that close in super-Earth planets (r<0.3 AU) are predominately rocky, but that there are also icy super Earths \citep{2013ApJ...775..105O, 2014ApJ...792....1L, 2018arXiv180501453F, 2018ApJ...853..163J, 2018MNRAS.479.4786V}. Only the models shown in Fig.~\ref{fig:icemignodisc} and Fig.~\ref{fig:icelowmigdisc} allow for rocky and ice rich inner planets within the same framework, but the water ice line of the model presented in Fig.~\ref{fig:icemignodisc} does not evolve in time, which we deem unrealistic.
  \label{tab:Results}
  }
 \end{table*}
 }%

\subsection{Migration}

In our models, we have tested three different migration prescriptions: (i) no migration (Fig.~\ref{fig:icenomignodisc} and Fig.~\ref{fig:icenomigdisc}), (ii) nominal outward migration (Fig.~\ref{fig:icemignodisc} and Fig.~\ref{fig:icemigdisc}) and (iii) only inward migration (Fig.~\ref{fig:icelowmigdisc}).

The radial extend in which outward migration is possible is related to the opacity transition at the water ice line \citep{2013A&A...549A.124B, 2014A&A...564A.135B, 2015A&A...575A..28B, 2015arXiv150303352B}. More precise, outward migration is possible exterior ($r>r_{\rm H_2O}$) to the water ice line. Planetary embryos forming in this region thus migrate outwards if they can reach the masses required for outward migration (this mass is higher in Fig.~\ref{fig:icemignodisc} compared to Fig.~\ref{fig:icemigdisc}). This will change the water ice fraction of the planets significantly. Additionally to that, pebble isolation happens at lower planetary masses compared to the maximum mass what the region of outward migration can contain \citep{2015A&A...582A.112B, Izidoro18}. Planets reaching pebble isolation mass exterior to the water ice line (up to a few AU beyond the water ice line) thus always end up in the region of outward migration. This thus results in a water ice fraction of $\sim$35\% of planets formed just exterior to the ice line (see Fig.~\ref{fig:icemigdisc}). The formed planets in our simulations seem to stay in the outer disc, however, mutual interactions can influence their migration and allow them to migrate inwards (see below).

This implies that if low mass planets are formed close to the water ice line and they can migrate outwards through the entropy related corotation torque then their water ice content is close to the maximum allowed by the chemical model. Planets formed in this way should thus have the same planetary composition and density, if their masses are similar. Early inward migration during the gas-disc phase on the other hand would also alter planetary densities, depending on the starting position of the embryo in respect to the ice lines. This additionally implies that a system like Trappist-1, where planets have slightly different densities (0.6-1.0$\rho_{\rm E}$ \citealt{2017Natur.542..456G}), is more likely to have formed through the inward migration of planets formed at the water ice line than through planets that initially migrate outwards after forming at the water ice line.

\subsection{Multiple growing embryos}

In our model we follow the growth of single planets, while in reality many planetary systems actually consist of multiple planets. This implies that the growing planets compete for solid material as they grow. In the pebble accretion scenario, the pebble flux is subsequently reduced for inner planets by the outer accreting planets \citep{2012A&A...544A..32L}. The inner planets thus have less material available and grow slower than the outer planets. However, this filtering is very inefficient for just one single planet, but as soon as many planets are present its effects become much stronger \citep{2014A&A...572A.107L}. Additionally, when the outer planets reach pebble isolation mass, they block the flux of pebbles to the inner systems and can starve the inner embryos quenching their growth \citep{2014A&A...572A..35L, 2015Icar..258..418M, 2018arXiv180102341B, Izidoro18, Bitsch18}.

In addition, the pebbles in the outer disc (at $r>r_{\rm H_2O}$) are supposed to be larger due to the larger fragmentation velocity of water ice grains compared to silicate grains \citep{2009ApJ...702.1490W, 2015ApJ...798...34G} as well as through condensation \citep{2013A&A...552A.137R, 2017A&A...602A..21S}. These larger pebbles are easier accreted and allow the planets to grow faster compared to the inner disc where pebbles are supposed to be only mm in size, corresponding to the size of chondrules. We have taken this effect into account in our simulations by reducing the growth timescale by a factor of 4, if no water ice is available. The faster growth in the outer disc can also explain why Jupiter in our own solar system grew so big compared to the terrestrial planets \citep{2015Icar..258..418M}. 

If multiple planetary embryos are present in a disc over a large radial extend in the disc spanning from interior to exterior to the water ice line, the planetary embryos exterior to the water ice line will grow faster and starve the inner planetary embryos \citep{Izidoro18, Bitsch18}. The then formed super-Earths are water rich in the \citet{Izidoro18} scenario due to outward migration that keeps the embryos at $r>r_{\rm H_2O}$ until they reach pebble isolation mass. Thus, it is of crucial importance where the first planetary embryos form: interior \citep{2011ApJ...728...20S, 2016A&A...596L...3I} or exterior \citep{2017A&A...608A..92D} to the water ice line as this then determines the final composition of the planets.

In the N-body simulations of \citet{Izidoro18} the planetary embryos forming exterior to the water ice line do not only grow faster, but they also migrate towards the inner edge of the protoplanetary disc, even though outward migration is possible. The reason why planets migrate towards the inner edge in multi-body simulations (see also \citealt{2017MNRAS.470.1750I} and \citet{2018MNRAS.479L..81R}) is that the mutual interactions between the planets increase their eccentricity which then quenches the entropy driven corotation torque \citep{2010A&A.523...A30} resulting in inward migration. Additionally, some of the growing planets reach masses (e.g. due to collisions) which are larger than the pebble isolation mass and also larger than the maximum mass the region of outward migration can contain and thus migrate inwards. Both effects are not taken into account in the here presented simulations.

On the other hand, if planets only migrate inwards, the exact formation location of the first embryos in respect to the water ice line might not matter that much as even embryos that are originally located at $r>r_{\rm H_2O}$ migrate across the water ice line and then accrete predominately rocky material, resulting in super-Earths with a low water ice content (Fig.~\ref{fig:icelowmigdisc}). 

The study by \citet{2018MNRAS.479L..81R} focuses on already formed planets of up to several Earth masses that migrate inwards from interior and exterior to the water ice line to form close-in super Earth systems. They find that the innermost super-Earth in a tightly packed system can be rocky, if it formed interior to the water ice line. Combining studies of growth of multiple embryos with different migration prescriptions might thus help to understand the formation pathway of super-Earths and will be investigated in a future study.

\subsection{Core heating by $^{26}$Al decay}

The early solar nebular was enriched with $^{26}$Al which decays to $^{26}$Mg with a halftime of 700kyr, releasing enough energy to evaporate water ice from pristine planetesimals in the solar system (e.g. \citealt{1993Sci...259..653G, 2018SSRv..214...39M}). Planets formed from these water poor planetesimals by mutual collisions would then also remain dry. This could apply also to planetary embryos formed exterior to the water ice line ($r>r_{\rm H_2O}$). For pebble accretion, this picture is slightly different. The pebbles are so small that heating by $^{26}$Al is inefficient and they would keep their water content. However, as the pebbles are accreted by the planet, they will also deliver some $^{26}$Al, which could then contributes to the total heating of the planet potentially allowing the water to evaporate. The pebble sizes on the other hand can determine the efficiency of the melting process of the planetary embryo in itself \citep{2019E&PSL.507..154L}.

On the other hand, if the planet forms on time scales comparable or longer than the radioactive decay time of $^{26}$Al, water rich pebbles will not deliver $^{26}$Al (because it already decayed) to the planet, so that the planet can become water rich by the accretion of water rich pebbles. This also applies to planets formed originally interior to the water ice line, but are then swept by the water ice line as the disc evolves.

This implies that if early formed planets can evaporate their water atmosphere through $^{26}$Al heating, they have to form fast enough while the effects of $^{26}$Al heating are still active. If the planets form too late, or accrete most of their mass after 1 Myr, $^{26}$Al heating might not prevent the planet from accreting a significant water content. On the other hand, internal heating of the planet itself after 1 Myr could evaporate some of the water and the atmosphere itself \citep{2018arXiv181103202G}. 

\citet{Lichtenberg19} showed that additionally the total amount of $^{26}$Al plays a crucial role in determining the planets fate if they are formed by planetesimal accretion. Low $^{26}$Al contents allow water rich planets to form, while a larger $^{26}$Al content only allows the formation of water poor planets. This clearly indicates that the process of $^{26}$Al heating in planet-formation simulations should be taken into account in future work. Our here presented simulations though show that different water contents of planets could be achieved instead by different migration prescriptions.

\subsection{Solar System formation}

The terrestrial planets in the solar system are extremely water poor, indicating that they formed in the hotter regions of the protoplanetary discs, where water ice was not available \citep{2018arXiv180308830I}. Earth's water, which matches the D/H and $^{15}$N/$^{14}$N ratios of carbonaceous chondrite meteorites \citep{2012Sci...337..721A} is thought to have been delivered by water-rich planetesimals that were gravitationally scattered into the inner Solar System either by asteroidal embryos \citep{2000M&PS...35.1309M, 2007AsBio...7...66R} or by Jupiter during its growth and/or migration \citep{2011Natur.475..206W, 2017Icar..297..134R, 2018SSRv..214...47O}.

The planetary embryos that formed the Earth were most likely quite small (about Mars mass), so that they have not migrated far and thus originated from a region around 1 AU. However, in most disc models \citep{2011ApJ...738..141O, 2015A&A...575A..28B, 2015arXiv150303352B}, the water ice line sweeps this region during the gas disc evolution. This would allow planetary embryos forming in this region to become water rich, even if they do not migrate (see Fig.~\ref{fig:icenomigdisc}). A solution to this problem was presented by \citet{2016Icar..267..368M}, who suggested that Jupiters core blocks the influx of water rich pebbles to the inner disc when it reaches pebble isolation mass thus starving the inner system. So even when the water ice line position would sweep the inner system, there would be no pebbles available to accrete for the interior embryos. The embryos in the inner disc thus only accrete rocky material at the beginning of their growth.

\subsection{Rocky super-Earths or waterworlds}

Detailed observations of planetary radii have revealed a gap in the radius distribution of transiting super-Earths at about 1.8 Earth radii \citep{2017AJ....154..109F, 2018arXiv180501453F}. Utilizing additionally RV measurements to determine the bulk density of the super-Earths allows to speculate about the origin in the planetary radius gap. The most popular theory to explain the radius gap is related to the photoevaporation of the planetary atmospheres by their host star in combination with a predominately rocky planetary composition \citep{2017ApJ...847...29O, 2018ApJ...853..163J}. This relation in itself is thus also build heavily on interior planet modelling, where also the long term evolution of the core itself can cause uncertainties up to 15\% in radius \citep{2018arXiv181102588V}.

Alternatively the gap could also be explained directly by cooling from the planet itself \citep{2018arXiv181103202G}. In their simulations, super-Earths could have water contents up to 20\%, implying that they formed exterior to the water ice line and then migrated inwards before their formation was complete, while the explanation due to photoevaporation requires only a very small water ice content inside the planets. This difference is probably caused by the differences in the interior modeling of the planets. 

If most close-in (r<0.3 AU) super-Earths are completely dry, then planetary embryos either only form interior to the snow line, which is at odds with the solar system and cold Jupiters in general. Alternatively it could imply that the water ice of pebbles or planetesimals evaporates as the planet forms \citep{2015IJAsB..14..201M}. This however, is also at odds with the solar system due to the icy nature of Uranus and Neptune. 

On the other hand, the model by \citet{2014A&A...562A..80K} implies that close-in super-Earths (with distances up to 0.03 AU) could additionally lose their whole water ice content through photoevaporation making them predominately rocky even if they were formed with a large water ice content as long as they do not exceed 3 Earth masses.

Additionally, the valley in the radius distribution observed by \citet{2017AJ....154..109F} and \citet{2018arXiv180501453F} is not empty and could also be shifted to larger planetary radii \citep{2018MNRAS.479.4786V}, implying that water rich super-Earths could exist to a significant fraction. In our model, the water ice content is determined by the migration and accretion of the planet and the water ice line evolution. As the disc becomes older, the water ice line moves interior to 1 AU at the end of the gas disc's lifetime. Late planet formation thus reduces the parameter space for the formation of purely rocky super-Earths, implying that the formation of purely rocky super-Earths probably happens early during the gas-disc phase, if planet migration is directed mostly inwards.

Hot super-Earths that form in-situ can {\it not} harbour any significant water ice content, because the water ice line does not evolve all the way down to 0.1 AU for reasonable parameters of disc models \citep{2007ApJ...654..606G, 2011ApJ...738..141O, 2015A&A...575A..28B, 2015arXiv150303352B}. In-situ formation is thus in contradiction to the observations, because the valley in the radius distribution is not empty, implying some super-Earths can harbour water. Only migrating planets that formed exterior to the water ice line at $r>r_{\rm H_2O}$ can form hot super-Earth with a significant water ice content, where the exact water ice content depends on the migration speed, accretion rate and water ice line evolution (Fig.~\ref{fig:icemigdisc} and Fig.~\ref{fig:icelowmigdisc}). This implies that systems of water rich hot super-Earths on low eccentricity orbits are signposts of inward planet migration during the gas-disc phase. This is for example the case for the planets in the Trappist-1 system, where the planets feature a water-ice content of 5-10\% \citep{2018A&A...613A..68G}.

It is thus important to measure precisely planetary radii and masses to determine the planet's bulk composition as it can give important clues about planet formation and the migration history of the system.

\section{Summary and Conclusions}
\label{sec:conclusions}

In this paper we present a disc evolution and planetary growth model that allows us to calculate the water ice fraction of formed planets. We have investigated an interplay between planetary migration, planetary accretion and ice line evolution on the composition of planets. Our main findings are summarized as follows (see also Table~\ref{tab:Results}):

\begin{itemize}
 \item[1)] If the water ice line does not evolve in time and if planets do not migrate, the water ice fraction of a planet is solely determined by its starting position in respect to the water ice line (Fig.~\ref{fig:icenomignodisc}). This scenario would imply that close-in super-Earths (r<0.3 AU) should all be rocky.
 
 \item[2)] If the water ice line does not evolve in time, but planets are allowed to migrate, water rich super-Earths can exist in the inner planetary systems, if migration is inwards. Only planets that form exterior to the water ice line ($r>r_{\rm H_2O}$) can contain water and their water ice content is then determined by a competition between growth and migration speed (Fig.~\ref{fig:icemignodisc}).
 
 \item[3)] If the water ice line evolves in time, but planets are formed in situ, their water ice content is determined by their relative position to the water ice line (Fig.~\ref{fig:icenomigdisc}). Even if planets are initially interior to the water ice line, they can contain a significant fraction of water ice, because of the inward movement of the water ice line \citep{2016A&A...589A..15S}. Slow accreting planets that are originally interior to the water ice line will thus have a larger water ice fraction than fast accreting planets originally placed interior to the water ice line, because the slow accreting planets accrete most of their material after the water ice line has swept across their orbit. A fast accretion of planets interior to the water ice line can thus prevent the formation of water rich planets as invoked in \citet{Izidoro18} to form rocky super-Earths.
 
 \item[4)] In disc models that allow outward migration due to the entropy related corotation torque, outward migration exists exterior to the water ice line ($r>r_{\rm H_2O}$) due to the transition in opacity at the water ice line \citep{2013A&A...549A.124B, 2014A&A...564A.135B, 2015A&A...575A..28B, 2015arXiv150303352B}. The pebble isolation mass to which planets accreting pebbles can grow, however, is always smaller than the maximum planetary masses that can undergo outward migration. Thus planetary embryos originating from exterior to the water ice line ($r>r_{\rm H_2O}$) accrete the maximum fraction of water ice possible in our model (Fig.~\ref{fig:icemigdisc}), similar to the planets in the model where planet migration is artificially turned off (Fig.~\ref{fig:icenomigdisc}). These fully formed water rich planets can then migrate inwards in systems of multiple bodies as the disc evolves and form systems of close-in water rich super-Earths \citep{2017MNRAS.470.1750I, 2018MNRAS.479L..81R, Izidoro18}.

 \item[5)] In the scenario where the ice line moves and planets migrate only inwards, the water ice content of planets that are originally placed interior to the water ice line is for most planets negligible, due to the faster inward migration of planets compared to the water ice line evolution. Planets formed originally exterior to the water ice line ($r>r_{\rm H_2O}$) can have different water ice content, depending on their initial position relative to the water ice line and their inward migration speed.
 
 \item[6)] Faster growth of planetary seeds can allow planets to become too massive to be contained in the region of outward migration at the end of disc dissipation. These planets stay at the inner edge of the disc and the fast growth model allows the formation of rocky massive inner super-Earths. However, the formation of water rich inner super-Earths is not possible. A slow planetary growth prevents rocky planets to form efficiently and they thus also do not migrate efficiently into the inner edge of the disc. Icy super-Earths are parked at in the region of outward migration as for nominal growth (Fig.~\ref{fig:fastslow}).

 \item[7)] As the water ice line does not evolve all the way to the central star, water rich hot super-Earths can not have formed in-situ and must have migrated or scattered inwards. Water rich super-Earths on low eccentricity orbits are thus a signpost of planet migration. In combination with the observations of planetary radius gap and its interpretation \citep{2017AJ....154..109F, 2018arXiv180501453F, 2017ApJ...847...29O, 2018ApJ...853..163J} which proposes that there are some icy super-Earths, this implies that inward planet migration during the gas-disc phase might be the norm for these super-Earths. The exact migration history (inward migration in chains of planets or inward migration as single planets during the early gas disc phase) can thus be told by the water ice content of the planet.
 
\end{itemize}

Our simulations indicate that the water ice content of hot super-Earths is a function of their migration speed and direction. Super-Earths with a high water content probably underwent outward migration close to the water ice line until they were fully formed before migrating inwards (in chains of resonance) or originate from far out in the discs, while super-Earths with a low water ice content formed close to the water ice line and migrated inwards during the early gas disc phase. The water ice content of hot super-Earths could thus reveal important information about the early migration history of the planet relative to the water ice line.

\begin{acknowledgements}

B.B., thanks the European Research Council (ERC Starting Grant 757448-PAMDORA) for their financial support. S.N.R. thanks the Agence Nationale pour la Recherche for support via grant ANR-13-BS05-0003-002 (grant MOJO). A. I. thank FAPESP for support via grants 16/19556-7 and 16/12686-2. We also thank an anonymous referee for her/his comments that helped to improve the manuscript.

\end{acknowledgements}

\bibliographystyle{aa}
\bibliography{Stellar}
\end{document}